\documentclass{birkmult}
 \newtheorem{thm}{Theorem}[section]
 
 \newtheorem{lem}[thm]{Lemma}
 
 \theoremstyle{definition}
 
 \theoremstyle{remark}

 \numberwithin{equation}{section}

\begin{document}
%
%
%
%
%
%
%
%
%

\title[Exponential Product Formulas]
{Results on Convergence in Norm of Exponential Product Formulas
and Pointwise of the Corresponding Integral Kernels}
\author[Takashi Ichinose]{Takashi Ichinose}
\address{%
Department of Mathematics, Kanazawa University, 
Kanazawa, 920--1192, Japan}
\email{ichinose@kenroku.kanazawa-u.ac.jp}


\author{Hideo Tamura}
\address{Department of Mathematics, Okayama University, 
Okayama, 700--8530, Japan}
\email{tamura@math.okayama-u.ac.jp}



\date{Feb., 2007}

\begin{abstract}
For the last one and a half decades it has been known that
the exponential product formula
holds also {\it in norm} in nontrivial cases. In this note, 
we review the results  on its convergence in norm
as well as pointwise of the integral kernels in the case 
for Schr\"odinger operators, with error bounds. Optimality 
of the error bounds is elaborated.
\end{abstract}

\maketitle
\section{Introduction}


The Trotter product formula, Trotter--Kato product formula 
or exponential product formula is usually a product formula 
which in strong operator topology approximates 
the group/semigroup with generator being a sum of two operators.
It is often a useful tool to study Schr\"odinger evolution
groups/semigroups in quantum mechanics and to study Gibbs 
semigroups in statistical mechanics. 

To think of a typical case, let 
$A$ and $B$ be selfadjoint operators in a Hilbert space ${\mathcal H}$
with domains $D[A]$ and $D[B]$ and $H := A+B$ their operator sum 
with domain $D[H] = D[A]\cap D[B]$. Assume that 
$H$ is selfadjoint or essentially selfadjoint on $D[H]$ and 
denote its closure
by the same $H$. Then Trotter \cite{Tr} proved
the unitary product formula  
\begin{eqnarray*}
&&[e^{-itB/2n}e^{-itA/n}e^{-itB/2n}]^n - e^{-itH} 
\rightarrow  0, \quad  \hbox{\rm strongly}, \\
&&\qquad\qquad
  [e^{-itA/n}e^{-itB/n}]^n - e^{-itH} 
 \rightarrow  0, \quad \hbox{\rm strongly},
\quad n\rightarrow \infty,
\end{eqnarray*} 
and also, when $A$ and $B$ are nonnegative, 
the selfadjoint product formula 
\begin{eqnarray*}
&& [e^{-tB/2n}e^{-tA/n}e^{-tB/2n}]^n - e^{-tH} 
\rightarrow  0,\quad  \hbox{\rm strongly}, \\
&&\qquad\quad\,\,\,\,\, [e^{-tA/n}e^{-tB/n}]^n - e^{-tH} 
\rightarrow  0,  \quad \hbox{\rm strongly}. 
\quad n\rightarrow \infty,
\end{eqnarray*}
The convergence is {\it locally uniform}, i.e.
uniform on compact $t$-intervals, respectively
in the real line ${\bf R}$ and in the closed half line $[0,\infty)$. 
Kato \cite{Ka} dicovered the latter selfadjoint product formula
to hold also for the form sum  $H : = A \dot{+}B$ with form domain 
$D[H^{1/2}] =D[A^{1/2}]\cap D[B^{1/2}]$, which we assume for simplicity
is dense in ${\mathcal H}$. However, it remains to be an open problem 
whether the unitary product formula for the form sum holds.

\medskip
However, since around 1993 we have begun to know that
selfadjoint product formulas converge even in ({\it operator}) {\it norm}, 
though in some special cases,
by the following two first results. Rogava \cite{Ro} proved, 
when $B$ is $A$-bounded and $H=A+B$ is selfadjoint, among others,
the abstract product formula that
$$\|[e^{-tA/n}e^{-tB/n}]^n - e^{-tH}\| =O(n^{-1/2}\log n),
\quad n\rightarrow \infty,
$$ 
locally uniformly in $[0,\infty)$.
Helffer \cite{He} proved, when $H := -\Delta +V(x)$ is
a Schr\"odinger operator in $L^2({\bf R}^d)$
with nonnnegative potential $V(x)$ satisfying
 $|\partial^{\alpha}_x V(x)|$ $\leq C_{\alpha}$ $ (|\alpha|\geq 2)$
so that $H$ is selfadjoint on the domain $D[-\Delta]\cap D[V]$, 
the symmetric product formula that
$$
 \|[e^{-tV/2n}e^{-t(-\Delta)/n}e^{-tV/2n}]^n - e^{-tH}\| = O(n^{-1}),
\quad n\rightarrow \infty,
$$
locally uniformly in $[0,\infty)$.
Many works were done to extend these results before 2000,
e.g. in
\cite{CZ1, IT1, IT3, NZ1, NZ2, NZ4} for the abstract product formula,  
\cite{DS, DIT, ITak1, ITak2, ITak3, Tak1} for the Schr\"odinger operators, 
and after that, e.g. in \cite{IT4, ITTZ, INZ}, \cite{CNZ1, CNZ2, CZ2, CZ3} 
for the abstract product formula. In most of them, use was made of 
operator-theoretic methods, though of a probabilistic method 
in \cite{ITak1, ITak2, ITak3, Tak1}.

In this note,  we want to describe more recent results
on convergence in norm for exponential product formulas
and also pointwise of the corresponding integral kernels,
mainly based on  our works since around 2000,
\cite{IT4, ITTZ, IT5, IT6, IT7}.
As for the error bounds, although it is easy to see 
by the Baker--Campbell--Hausdorff formula (e.g. \cite{V}, \cite{Su}) 
that with both operators $A$ and $B$ being bounded, 
the {\it nonsymmetric} product formula has an optimal error bound
$O(n^{-1})$ while the {\it symmetric} one does $O(n^{-2})$, 
it was shown in \cite{ITTZ} that even the {\it symmetric} product formula
has an optimal error bound $O(n^{-1})$ in general, 
if both $A$ and $B$ are unbounded. 
However, in \cite{IT6} (cf. \cite{IT7}), a better upper sharp error 
bound $O(n^{-2})$ has been obtained for 
the {\it symmetric} product formula
with the Schr\"odinger operator $-\Delta +V(x)$ having
nonnegative potentials $V(x)$ growing polynomially at infinity,
in spite that both $-\Delta$ and $V$ are unbounded operators. 
In this note we mention, with a sketch of proof,
a latest complementary result \cite{AI} which settles the sharp 
optimal error bound is in fact $O(n^{-2})$ 
with the symmetric product formula for the harmonic oscillator,
by estimating the error not only from 
above but also from below, in norm as well as pointwise.

Theorems are described in Section 2. Optimality of error bounds 
is discussed separately in Section 3. The idea of proof 
is briefly mentioned in Section 4.  In Section 5 we give concluding 
remarks, and also refer to a connection of the exponential product 
formula with the Feynman path integral.

It should be also noted that in almost the same context with the
notion of norm ideals (e.g. \cite{GK}, \cite{S}) we are able 
to deal with the trace norm convergence as  in
\cite{Za1, NZ0a, NZ0b, NZ3, Hi, IT2, Tak2}. 
For an extensive literature on this we refer to \cite{Za2}.

\bigskip
The content of this note is an expanded version of the lecture 
entitled ``On converegence pointwise of integral kernels and 
in norm for exponential product formulas" given by T.I. 
at the International Conference ``{\sl Modern Analysis 
and Applications} (MAA 2007)", Odessa, Ukraine, 
April 9--14, 2007, which is a slightly extended version of 
of the lecture (unpublished) given at the Conference on
``{\sl Heat Kernel in Mathematical Physics}",
Blaubeuren, Germany, November 28--December 2, 2006.


\section{Theorems}

We begin with our result which extends ultimately 
Rogava and Helffer's.
 
\begin{thm}\label{TH1} 
(Ichinose-Tamura-Tamura-Zagrebnov  2001\cite{IT4, ITTZ})$\,\,$
Let $A$ and $B$ be nonnegative selfadjoint operators, and assume
$H =A+B$ is selfadjoint on $D[H]=D[A]\cap D[B]$. 
Then as $n\rightarrow\infty$,
\begin{eqnarray}
 \|[e^{-tB/2n}e^{-tA/n}e^{-tB/2n}]^n - e^{-tH}\| &=& O(n^{-1}),
                \label{operator-Trotter1}\\
 \|[e^{-tA/n}e^{-tB/n}]^n - e^{-tH}\| &=&  O(n^{-1}).
                \label{operator-Trotter2}
\end{eqnarray}
The convergence is locally uniform in the closed half line $[0,\infty)$,
while on the whole half line $[0,\infty)$, if $H$ is strictly positive,
i.e. $H \geq \eta I$ for some $\eta>0$.
The error bound $O(n^{-1})$ in (\ref{operator-Trotter1}) and 
(\ref{operator-Trotter1}) is optimal.
\end{thm}


We can go beyond this result. 
First, focussing on the Schr\"odinger operator $-\Delta + V(x)$,
we ask whether norm convergence implies pointwise convergence 
of integral kernels. The answer is yes, though strong convergence does not.
This problem is discussed for 
 Schr\"odinger operators with potentials of polynomial growth 
(Theorem~\ref{TH2}),
 with positive Coulomb potential (Theorem~\ref{TH3}), 
and also for the Dirichlet Laplacian (Theorem~\ref{TH4}).
Pointwise convergence of integral
kernels for Schr\"odinger semigroups is important, because
it gives a  time-sliced approximation to the imaginary-time 
Feynman path integral.

Next, we ask, for the unitary exponential/Trotter product formula,
 whether there are nontrivial cases where it converges in norm,
though it does not in general hold (see \cite{I}).
The answer is yes. In fact, it holds for the Dirac operator 
and relativistic Schr\"odinger operator (Theorem~\ref{TH5}).


Let $H = H_0+V := -\Delta +V(x)$ with $V(x)$ a real-valued function.
By $K^{(n)}(t,x,y)$ we denote the integral kernel of
$[e^{-tH_0/2n}e^{-tV/n}e^{-tH_0/2n}]^n$, and by $e^{-tH}(x,y)$
that of $e^{-tH}$.

\begin{thm}\label{TH2}(Ichinose-Tamura 2004 \cite{IT6})
(positive potential of polynomial growth) $\,\,$ 
 Assume that $V(x)$ is in $C^{\infty}({\bf R}^d)$, bounded below  
and satisfies
$|\partial_x^{\alpha}V(x)| \leq C_{\alpha} 
\langle x\rangle^{m-\delta|\alpha|}$ with some $0< \delta \leq 1$
($\langle x\rangle =(1+x^2)^{1/2}$).

\noindent
(i) (In norm)
\begin{equation}
\|[e^{-tH_0/2n}e^{-tV/n}e^{-tH_0/2n}]^n-e^{-tH}\|_{L^2} 
 = O(n^{-2}),
\end{equation}
locally uniformly in the open half line $(0,\infty)$.

\medskip\noindent
(ii) (Integral kernel) $\,\,$
\begin{eqnarray}
  &&[K^{(n)}(t,x,y) -e^{-tH}(x,y)] = O(n^{-2}), \nonumber\\
  &&
 \hbox{\rm in}\,\,\, 
   C^{\infty}({\bf R}^d \times {\bf R}^d)\hbox{\rm -topology},\,\,
  \hbox{\rm locally uniformly}\,\,\hbox{\rm in}\,\, (0,\infty) ,
\end{eqnarray}
i.e. together with all $x,\, y$-derivatives. 
\end{thm}

This theorem improves the result of Takanobu \cite{Tak1}, who used
a probabilistic method with the Feynman--Kac formula (see Sect. 5)
to show uniform pointwise convergence of the integral kernels, 
roughly speaking, 
with error bound $O(n^{-\rho/2})$, if $V(x)$ satisfies
$V(x)\geq C (1+|x|^2)^{\rho/2}$ and $|\partial_x^{\alpha}V(x)|$ 
$\leq C_{\alpha} (1+|x|^2)^{(\rho-\delta|\alpha|)_+/2}$
for some constants $C,\, C_{\alpha} \geq 0$ and 
$\rho \geq 0,\,0<\delta \leq 1$. 
The claim of Theorem 2.2 is a little bit sharpened 
in Theorems 3.1 and 3.2, in the next section,  
in the case of the harmonic oscillator.

\begin{thm}\label{TH3}
(Ichinose-Tamura 2006 \cite{IT7}) (positive Coulomb potential)$\,\,$ 
Let $H := -\Delta + V(x)$ with $V(x) \geq 0$.
Assume  that
$V(-\Delta +1)^{-\alpha}$:
 $L^2({\bf R}^d) \rightarrow L^2({\bf R}^d)$ is bounded   
   for some $0< \alpha <1$,
and that
$V \in C^{\infty}$ near a neighbourhood $U$ of both $p$ and $q$ 
(after $p, q \,\,\in {\bf R}^d$ taken). 
Then
\begin{eqnarray}
  &&[K(t/n)^n(x,y) -e^{-tH}(x,y)] = O(n^{-1}), \nonumber\\
  &&  \hbox{\rm in}\,\, C^{\infty}(U)\hbox{\rm -topology},\,\,
  \hbox{\rm locally uniformly}\,\,\hbox{\rm in}\,\, (0,\infty) .
\end{eqnarray}

\end{thm}

The condition is satisfied if
$V$ is in $L^2({\bf R}^3) + L^{\infty}({\bf R}^3)$, in particular,
if $V$ is the positive Coulomb potential $1/|x|$.
We don't know what happens at the singularities of $V(x)$.

\begin{thm}\label{TH4} (Ichinose-Tamura 2006 \cite{IT7})
(Dirichlet Laplacian)$\,\,$
Let $\Omega \subset {\bf R}^d$ be a bounded domain with smooth boundary
and $\chi_{\Omega}$ the indicator function of $\Omega$.
Let
$H_0 := -\Delta$ in $L^2({\bf R}^d)$, and  
$H := -\Delta_D$ the Dirichlet Laplacian in $\Omega$
with domain $D[H] = H^2(\Omega) \cap H_0^1(\Omega)$. 
Then for $0< \sigma < \frac16$,
\begin{eqnarray}
&&\big(\chi_{\Omega}\, e^{-tH_0/n}\, \chi_{\Omega}\big)^n(x,y) -e^{-tH}(x,y)
= O(n^{-{\sigma}}), \nonumber\\
&&\hbox{\rm locally uniformly}\,\,\hbox{\rm in}\,\,(t, x,y) \in
  (0,\infty) \times{\Omega} \times{\Omega}.
\end{eqnarray}

\end{thm}

We don't know what happens when $x$ or $y$ approaches the boundary
of $\Omega$.

\medskip\noindent
{\bf{Corollary}.}
$$ 
\|[\chi_{\Omega}\,e^{-tH_0/n}\,\chi_{\Omega}]^n f-e^{-tH}f\|_{L^2}
 \rightarrow 0, \quad f \in L^2(\Omega).
$$

\bigskip
Consequently, Theorem~\ref{TH4} is a {\it stronger} statement 
than this corollary,
though the latter is also obtained by Kato \cite{Ka} as
an abstract result: If $A$ is a nonnegative selfadjoint operator
and $P$ an orthogonal projection in a
Hilbert space ${\mathcal H}$, then
$(Pe^{-tA/n}P)^n \rightarrow e^{-tA_P}, \,\, strongly,
$
as $n\rightarrow \infty$,
where $A_P := (A^{1/2}P)^*(A^{1/2}P)$. In passing, however, it is 
an open question whether it holds that
$(Pe^{-itA/n}P)^n \rightarrow e^{-itA_P}P$, $\,$ strongly
({\it Zeno product formula}). A partial answer 
was given in \cite{EI}.

All Theorems 2.2--2.4 hold with order of products
exchanged,  e.g.  in Theorem~\ref{TH2},
 $[e^{-tV/2n}e^{-tH_0/n}e^{-tV/2n}]^n$
instead of $[e^{-tH_0/2n}e^{-tV/n}e^{-tH_0/2n}]^n$.


\begin{thm}\label{TH5} 
(Ichinose-Tamura 2004 \cite{IT5})(Unitary Trotter in norm)$\,\,$ 
Let $A$ and $B$ be selfadjoint, and assume $H :=A+B$ to be 
essentially selfadjoint in a Hilbert space ${\mathcal H}$.
Assume that there exists a dense subspace ${\mathcal D}$ of ${\mathcal H}$
with ${\mathcal D} \subset D[A]\cap D[B]$ such that 
$e^{-itA},\, e^{-itB}: {\mathcal D} \rightarrow {\mathcal D}$. 
Further assume that the commutators $[A,B]$, $[A, [A,B]]$ and 
$[B,[A,B]]$ are  bounded on ${\mathcal H}$.
Then
\begin{equation}
 \|\bigl(e^{-itB/2n}e^{-itA/n}e^{-itB/2n}\bigr)^n - e^{-itH}\| =
 O( n^{-2}),  \quad n\rightarrow \infty,
\end{equation}
locally uniformly in the real line ${\bf R}$.
\end{thm}

As important applications we have ones to the Dirac operator 
$H= H_0+V= (i {\alpha}\cdot \nabla + m\beta) +V(x)$ in $L^2({\bf R}^3)^4$,
where $\alpha =(\alpha_1, \alpha_2, \alpha_3)$ and $\beta$
are the 4 Dirac matrices, 
with  $\partial^{\gamma}V(x)(|\gamma|=2)$ being bounded,
as well as to the relativistic Schr\"odinger operator
$H=H_0+V= \sqrt{-\Delta +m^2} +V(x)$ on $L^2({\bf R}^d)$
with $\partial_x^{\gamma} V(x)$ being bounded  
for $1 \leq |\gamma| \leq 4$ ($0 \leq |\gamma| \leq 4$, if  $m=0$).
In these cases, $H$ are essentially selfadjoint, and satisfy
the conditions in the theorem. So it holds that
\begin{equation}
\|[e^{-itV/2n}e^{-itH_0/n}e^{-itV/2n}]^n-e^{-itH}\|_{L^2} 
 = O(n^{-2}),  \quad n\rightarrow \infty,
\end{equation}
locally uniformly in ${\bf R}$.

\smallskip
However, this theorem does not apply to Schr\"odinger operators
except for the {\it Stark Hamiltonian} 
$(-\Delta + V(x))+ a\cdot x$ in $L^2({\bf R}^d)$, 
where $a$ is a constant real vector in ${\bf R}^d$.

Finally it should be noted that we have shown in 
Theorems 2.2--2.4 that
the convergence is unifrom only on compact $t$-intervals
which are away from $0$, though in Theorems~\ref{TH1} and \ref{TH5},  
on ones which are allowed to be not away from $0$.


\section{Optimalty of Error Bounds}

In this section we discuss optimality of error bounds.
The error bound $O(1/n)$ in Theorem~\ref{TH1} is optimal, because 
if both $A$ and $B$ are bounded operators, by the 
Baker--Campbell--Hausdorff formula we know
\begin{eqnarray*}
 &&[e^{-tA/n}e^{-tB/n}]^n - e^{-tH} =  R'_n \cdot  n^{-1},  \\
 &&[e^{-tB/2n}e^{-tA/n}e^{-tB/2n}]^n - e^{-tH} =  R_n \cdot n^{-2}, 
\end{eqnarray*}
for some $ R'_n$ and $R_n$ being uniformly bounded operators which 
in general are not the zero operator.
From this,  optimality in the former non-symmetric case is evident.
But even in the symmetric case it is optimal. Indeed, 
there exist unbounded nonnegative selfadjoint operators  $A,\,B$ 
such that $H=A+B$ is selfadjoint and
$$\|[e^{-tB/2n}e^{-tA/n}e^{-tB/2n}]^n - e^{-tH}\|
\geq c(t) n^{-1}
$$
for some continuous function $c(t)$ with $c(t)>0,\,\, t>0$ 
and $c(0)=0$ (\cite{ITTZ}).

However, further in some special symmetric case in Theorem~\ref{TH2} where 
$-\Delta,\, V$ are taken as $A,\, B$, we have seen  
the symmetric product formula hold with a {\it sharp} error bound
$O(n^{-2})$.
We can make more precise this result with
the 1-dimensional harmonic oscillator
$H := H_0+V := \frac12 (-\partial_x^2 +x^2)$ in $L^2({\bf R})$.

 
\begin{thm}\label{TH6}(Azuma-Ichinose 2007 \cite{AI})$\,\,$
There exists bounded continuous functions 
$C(t) \geq 0$ and $c(t)\geq 0$  in $t\geq 0$, which are 
positive except $t=0$ with $C(0)=c(0)=0$, 
independent of $n$, such that for $n= 1, 2, \dots$,
\begin{equation} 
 c(t)n^{-2} \leq 
\|[e^{-\frac{t}{2n}V}e^{-\frac{t}{n}H_0}e^{-\frac{t}{2n}V}]^n
    - e^{-tH}\| \leq C(t) n^{-2}, \qquad t \geq 0\,. 
\end{equation}

\end{thm}

This theorem mentions  an error bound from below,
extending the harmonic oscillator case of 
Theorem~\ref{TH2} which treats only the right-half inequality 
with $C(t) =C$ being a positive constant depending 
on each compact $t$-interval in the {\it open} half line $(0,\infty)$.  

It is anticipated that the same is true for the Schr\"odinger operator
$H =-\Delta + V(x)$ with growing potentials like 
$V(x) = |x|^{2m}$ treated in Theorem~\ref{TH2}.

\medskip
Theorem~\ref{TH6} is obtained as a corollary from the following theorem
of its integral kernel version. 
Here one calculates explicitly the integral kernel
$K^{(n)}(t,x,y)$ of $[e^{-tV/2n}e^{-tH_0/n}e^{-tV/2n}]^n$ 
to estimate its difference from the integral kernel 
$e^{-tH}(x,y)$ of $e^{-tH}$.

\begin{thm}\label{TH7}(Azuma-Ichinose 2007 \cite{AI})$\,\,$
There exists a bounded operator $R(t)$ and uniformly bounded
operators $\{Q^{(n)}(t)\}_{n=1}^{\infty}$ 
with integral kernels $R(t,x,y)$ and $Q^{(n)}(t,x,y)$
being  uniformy bounded continuous functions 
in $(0,\infty)\times {\bf R}\times {\bf R}$ such that
\begin{equation}
\!\! K^{(n)}(t,x,y)- e^{-tH}(x,y) = 
 \big[R(t,x,y) + Q^{(n)}(t,x,y)n^{-1}\big] n^{-2};
\end{equation}
they satisfy
$$ 
 \sup_{x,y}|R(t,x,y)|,\, 
 \sup_n\sup_{x,y}|Q^{(n)}(t,x,y)| \rightarrow 0, \,\, t\rightarrow 0 \,\,; 
 \quad \sup_{x,y}|R(t,x,y)| \rightarrow 0, \,\, t\rightarrow \infty.
$$
$R(t,x,y)$ is explicitly given by
\begin{eqnarray}
R(t,x,y) 
&\!=\!& e^{-tH}(x,y)\frac{t^2}{12}\Big[
t\Big(\frac1{4}
    \frac{e^t+e^{-t}}{e^t-e^{-t}}
     +\frac{(e^t+e^{-t})xy-(x^2+y^2)}{(e^t-e^{-t})^2}\Big) \nonumber\\
&&\qquad\qquad\qquad\quad +\frac1{16}\Big(1+ 
   \frac{4xy-(e^t+e^{-t})(x^2+y^2)}{e^t-e^{-t}}\Big)\Big].
\end{eqnarray}
If $t>0$, $R(t,x,y)$ can become positive and negative. 
\end{thm}


\begin{lem} $K^{(n)}(t,x,y)=$
\begin{eqnarray}
&\!\!& \frac{1}{\sqrt{\pi}}\Biggl(\frac{\sqrt{1+\frac{t^2}{4n^2}}}
  {\Big(1+\frac{t}{n}\sqrt{1+\frac{t^2}{4n^2}}+\frac{t^2}{2n^2}\Big)^n
-\Big(1-\frac{t}{n}\sqrt{1+\frac{t^2}{4n^2}}+\frac{t^2}{2n^2}\Big)^n}
\Biggr)^{1/2}      \nonumber\\
&&\times 
\exp{\Biggl[\frac{2\sqrt{1+\frac{t^2}{4n^2}}}
{\Big(1+\frac{t}{n}\sqrt{1+\frac{t^2}{4n^2}}+\frac{t^2}{2n^2}\Big)^n
-\Big(1-\frac{t}{n}\sqrt{1+\frac{t^2}{4n^2}}+\frac{t^2}{2n^2}\Big)^n}}xy
\Biggr]             \label{3rdexpresprodkern-n} \nonumber\\
&&\times 
\exp\Biggl\{\Biggl[-\frac{t}{4n}-\frac{n}{2t} \Biggl(1- \nonumber\\
&& 
\frac{\Big(1+\frac{t}{n}\sqrt{1+\frac{t^2}{4n^2}}+\frac{t^2}{2n^2}\Big)^{n-1}
-\Big(1-\frac{t}{n}\sqrt{1+\frac{t^2}{4n^2}}+\frac{t^2}{2n^2}\Big)^{n-1}}
{\Big(1+\frac{t}{n}\sqrt{1+\frac{t^2}{4n^2}}+\frac{t^2}{2n^2}\Big)^n
-\Big(1-\frac{t}{n}\sqrt{1+\frac{t^2}{4n^2}}+\frac{t^2}{2n^2}\Big)^n}
  \Biggr)\Biggr](x^2+y^2)\Biggr\}. \nonumber\\
&& \label{kernel-diff-formula}
\end{eqnarray}
\end{lem}

{\it Proof}. Calculate the Gaussian integral
\begin{eqnarray*}
&&K^{(n)}(t,x,y)   \\
&\equiv& \Big(\frac{n}{2\pi t}\Bigr)^{\frac{n}{2}}
        \overbrace{\int_{\bf{R}}\cdots\int_{\bf{R}}}^{\mbox{$n-1$}}
        \prod_{j=1}^n \Big[e^{-\frac{t}{4n}x_j^2}
        e^{-\frac{(x_j-x_{j-1})^2}{2t/n}}
        e^{-\frac{t}{4n}x_{j-1}^2}\Big]
        \,dx_1\cdots dx_{n-1},
\end{eqnarray*}
where $x=x_n,\,\, y=x_0$.
We shall encounter with continued fraction to lead to the final expression
(\ref{kernel-diff-formula}) of the lemma.

\bigskip
To show Theorem~\ref{TH7}, 
we simply calculate the difference $K^{(n)}(t,x,y)- e^{-tH}(x,y)$,
though it is not so simple.

\smallskip
Here we mention what the operator with $R(t,x,y)$ as its
 integral kernel is. By the Baker--Campbell--Hausdorff formula
(e.g. \cite{V}, \cite{Su}),
if $A$ and $B$ are bounded operators, we have
\begin{eqnarray*}
&&\!\!\!\! [e^{-tB/2n}e^{-tA/n}e^{-tB/2n}]^n -e^{-t(A+B)} \\
&\!\!=\!\!& \exp \big(-t(A+B) 
      - n^{-2} \frac{t^2}{24}[2A+B,[A,B]] - O_p(n^{-3})\big)\\
&\!\!=\!\!& e^{-t(A+B)}-n^{-2} \frac{t^2}{24}
        \int_0^t e^{-(t-s)(A+B)}[2A+B,[A,B]]e^{-s(A+B)} ds
            + O_p(n^{-3}), 
\end{eqnarray*}
where $O_p(n^{-3})$ is an operator with norm of $O(n^{-3})$.
In our case where $A= -\frac12\partial_x^2,\,\, B= \frac12x^2$, 
we can show $R(t,x,y)$ is just the integral kernel of the operator
$$
   -\frac{t^2}{24}
        \int_0^t e^{-(t-s)H}[2H_0+V,[H_0,V]]e^{-sH} ds,
$$
which {\it does} make sense, though $H_0$ and $V$ are unbounded operators.
We have $[2H_0+V,[H_0,V]] = -4H_0+2V=-4H+6V$.


\section{Idea of Proof}
Put 
$K(\tau) = e^{-\tau B/2}e^{-\tau A}e^{-\tau B/2}$. Note that
$0 \leq K(\tau) \leq 1$. 
Then we need to estimate the difference between $K(t/n)^n$ and $e^{-tH}$.
The general technique of proof is: 
(i) to establish an appropriate version of Chernoff's theorem
(\cite{C}):
\begin{eqnarray*}
&&\big[(1+\tau^{-1}(1-K(\tau))^{-1}-(1+H)^{-1}\big] \rightarrow 0,
\,\,\,\tau\downarrow 0\qquad\qquad\qquad\qquad\qquad\qquad \\
 &&\qquad\qquad\qquad\qquad\qquad \Longrightarrow \quad
  [K({t}/{n})^n -e^{-tH}] \rightarrow 0, \,\,\, n\rightarrow\infty
\end{eqnarray*}
and/or (ii) to do telescoping:
\begin{eqnarray*}
 &&e^{-tH} -K(t/n)^n = \sum_{k=1}^ne^{-(k-1)tH/n}
                       (e^{-tH/n}-K(t/n))K(t/n)^{n-k}\\
\end{eqnarray*}
to estimate each summand on the right.
The former method (i) seems to be more efficient than the latter (ii). 

In fact, to prove Theorem 1, we use the former method, establishing
the following norm version  of Chernoff's  theorem with error bounds. 
The case without error bounds was noted by Neidhardt--Zagrebnov \cite{NZ2}.

\begin{lem} (Ichinose-Tamura \cite{IT4})
I.  Let $\{F (t)\}_{t \geq0}$ be a 
family of selfadjoint operators with $0 \leq F (t) \leq 1$,
and $H \geq 0$ a selfadjoint operator in a Hilbert space ${\mathcal H}$. 
Define $S_t := t^{-1} (1 - F (t))$.
Then 

\noindent
$\displaystyle{
\hbox{\rm{(a)}}\quad \hbox{\rm{For}}\,\, 0 < \alpha \leq 1,
\,\,\,
\|(1 + S_t)^{-1} - (1 + H)^{-1}\| = O(t^{\alpha}), 
\quad t \downarrow 0}$

\noindent
implies

\noindent
$\displaystyle{
 \hbox{\rm{(b)}}\quad \hbox{\rm{For}}\,\, \hbox{\rm{every}}\,\,
\hbox{\rm{fixed}}\,\, \delta > 0,}$

$\displaystyle{
\|F (t/n)^n - e^{-tH}\| 
= \delta^{-2}t^{-1 + \alpha} e^{\delta t}O(n^{-\alpha}), 
   \,\,\, n \rightarrow \infty,  \,\,\, t>0.}
$ 

\smallskip
Therefore for $\alpha=1$ this convergence is uniform on each compact 
interval $[0,L]$ in the closed half line $[0, \infty)$.

\smallskip\noindent
II. Moreover, in case $H \geq \eta I$ for some constant $\eta > 0$, 
if for every $\varepsilon >0$ there exists 
$\delta(\varepsilon) >0$ such that 
$F(t) \leq 1-\delta(\varepsilon)$ for all $t \geq \varepsilon$, then
$$
\|F (t/n)^n - e^{-tH}\| 
= (1+ 2/\eta)^2 t^{-1+ \alpha} O(n^{-\alpha}), 
   \quad n \rightarrow \infty,\,\,\, t>0.
$$

Therefore for $\alpha =1$ this convergence is uniform on the 
whole closed half line $[0, \infty)$.
\end{lem}

\smallskip
Condition II is satisfied,  e.g. for 
$F(\tau) = e^{-\tau B/2}e^{-\tau A}e^{-\tau B/2}$.
For the proof, we refer to \cite{IT4}.

\bigskip
For the proof of Theorems 2.2--2.5 we employ 
the latter method (ii), and further, for Theorems 2.2--2.4, 
make a crucial use of Agmon's kernel theorem:

\begin{lem}(Agmon's kernel theorem \cite{Ag})
 
Let
$T: L^2({\bf R}^d) \rightarrow L^2({\bf R}^d)$ be a bounded operator 
with ranges of $T$ and its adjoint $T^*$ satisfying
$R[T], \,\, R[T^*] \subset 
H^m({\bf R}^d), \,\,m>d$.
If $T$ is an integral operator with integral kernel $T(x,y)$ 
being a bounded continuous function in
${\bf R}^d \times {\bf R}^d$ such that
$$(Tf)(x) = \int T(x,y) f(y)\, dy, \qquad f \in L^2,
$$
then
$$
|T(x,y)| \leq 
 C(\|T\|_m + \|T^*\|_m)^{\frac{d}{m}}\|T\|^{1-\frac{d}{m}},\,\,\,\,
$$
where $\|T\|_m := \|T\|_{{\mathcal L}(L^2\rightarrow H^m)}$ is the
operator norm of $T$ as a bounded operator of $L^2({\bf R}^d)$
into the Sobolev space $H^m({\bf R}^d)$.
\end{lem}

Indeed, we estimate the ${\mathcal L}(L^2\rightarrow H^m)$-operator norm
of the difference 
$T=
[e^{-tV/2n}e^{t\Delta/n}e^{-tV/2n}]^n -e^{t(-\Delta +V)}$.



\section{Concluding Remarks}

We have so far considered the case where
the {\it operator sum} $H =A+B$ of two nonnegative selfadjoint 
operators $A$ and $B$ is selfadjoint. However, otherwise, 
the exponential product formula in norm does not in general hold
for the {\it form sum} $H =A+B$ of two selfadjoint operators $A\geq 0$, 
$B\geq 0$, even if it is essentially selfadjoint on $D[A]\cap D[B]$
(see \cite{Tam}).
Nevertheless, there is some case where it holds:

\begin{thm}\label{TH8}
(Ichinose-Neidhardt-Zagrebnov 2004 \cite{INZ})$\,\,$
Let $H = A \dot{+} B$ be the form sum of $A$ and $B$.
If $D[H^{\alpha}] \subseteq D[A^{\alpha}] \cap D[B^{\alpha}]$
for some $\frac12 < \alpha<1$, and 
$D[A^{\frac12}] \subseteq D[B^{\frac12}] $, 
then
\begin{eqnarray}
  \|[e^{-tB/2n}e^{-tA/n}e^{-tB/2n}]^n - e^{-tH}\| 
   &\!=\!& O(n^{-(2\alpha -1)}), \label{form-Trotter1}\\
  \|[e^{-tA/n}e^{-tB/n}]^n - e^{-tH}\| 
   &\!=\!& O(n^{-(2\alpha -1)}), \label{form-Trotter2}
\end{eqnarray}
locally uniformly in $[0,\infty)$.  

\end{thm}  

This error bound in (\ref{form-Trotter1})/(\ref{form-Trotter2})  
is also optimal. For this we refer to \cite{Tam}.
The condition for the domains of $A$ and $B$ is not symmetric.
It is an open question whether one may improve it so as to become
symmetric with respect to $A$ and $B$.

\bigskip
Finally, as we should like to mention,
there is a very nice Feynman path integral formula
which represents the Schr\"odinger semigroup,
 called the {\it Feynman--Kac formula} 
\begin{eqnarray*}
 &&(e^{-tH}f)(x)
 =(e^{-t(-\Delta+V)}f)(x)\\
 &=&
 \int_{B \in C([0,\infty)\rightarrow{\bf R}^d), B(0)=x}
 \exp[-\int_0^tV(B(s))ds] f(B(t)) d \mu(B),
\end{eqnarray*}
where $\mu(\cdot)$ is the Wiener measure on the path space
$C([0,\infty)\rightarrow{\bf R}^d)$ (e.g. (\cite{Si}). 
We may use this formula to get {whatever} results, in fact, 
a lot of them. This is a {big advantage}!
But {disadvantage} is
 that it is only restricted to the Schr\"odinger operator 
or Laplacian. For instance, if we think of the semigroup
for the relativistic Schr\"odinger operator
$H= \sqrt{-\Delta +m^2}+V(x)$,
we have to establish another Feynman--Kac formula (cf. \cite{ITam}).

Indeed, the Feynman--Kac formula is one of the realizations 
of Feynman path integral as a {\it true integral} on a path space.
However, as Nelson \cite{Ne} noted, the exponential/Trotter product formula 
also can give a meaning to the Feynman path integral as 
a {\it time-sliced approximation} 
by finite-dimensional integrals (cf. \cite{I}).
What it has advantage at is that we may apply it to the 
sum $H=A+B$ of any two selfadjoint operators $A,B$
bounded from below.

\bigskip\noindent
{\it Acknowledgements}.
One of the authors (T.I.) would like to thank 
Professor Vadim Adamyan for kind and warm 
hospitality during his stay in Odessa for the Mark Krein 
Centenary Conference, April 2007.
This work (of T.I. and H.T.) was partially supported by Grant-in-Aid for 
Exploratory Research No. 17654030 
and by Grant-in-Aid for Scientific Research (B) No. 18340049, 
Japan Society for the Pomotion of Sciences.




\begin{thebibliography}{1}

\bibitem{Ag}
 S.~Agmon,
\emph{On kernels, eigenvalues, and eigenfunctions of operators
related to elliptic problems},
Comm. Pure Appl. Math. {\bf 18} (1965), 627--663.

\bibitem{AI}
 Y.~Azuma and T.~Ichinose,
\emph{Note on norm and pointwise convergence of exponential 
products and their integral kernels for the harmonic oscillator}, 
to appear in Integral Equations Operator Theory 
(published on line first Nov. 12, 2007).

\bibitem{CNZ1} V.~Cachia, H.~Neidhardt and V.~A.~Zagrebnov,
\emph{Accretive perturbations and error estimates for the Trotter 
product formula}, 
Integral Equations Operator Theory {\bf 39} (2001), 396--412. 

\bibitem{CNZ2} V.~Cachia, H.~Neidhardt and V.~A.~Zagrebnov,
\emph{Comments on the Trotter product formula error-bound estimates 
for nonself-adjoint semigroups}, 
Integral Equations Operator Theory {\bf 42} (2002), 425--448. 


\bibitem{CZ1} V.~Cachia and V.~A.~Zagrebnov,
\emph{Operator-norm convergence of the Trotter product formula for 
sectorial generators}, Lett. Math. Phys. {\bf 50} (1999), 203--211. 

\bibitem{CZ2} V.~Cachia and V.~A.~Zagrebnov,
\emph{Trotter product formula for nonself-adjoint Gibbs semigroups},
J. London Math. Soc. (2) {\bf 64} (2001), 436--444. 

\bibitem{CZ3} V.~Cachia and V.~A.~Zagrebnov,
\emph{Operator-norm convergence of the Trotter product formula 
for holomorphic semigroups}, 
 J. Operator Theory {\bf 46} (2001), 199--213. 

\bibitem{C} P.R.~Chernoff,
Product Formulas, Nonlinear Semigroups, and Addition of
Unbounded Operators,
Memoirs Amer. Math. Soc., {\bf 140}, 1974.

\bibitem{DS} B.~O.~Dia and M.~Schatzman, 
\emph{An estimate on the Kac transfer operator},
J. Functional Analysis {\bf 145} (1997), 108--135.

\bibitem{DIT} A.~Doumeki, T.~Ichinose and H.~Tamura,
\emph{Error bounds on exponential product formulas for 
Schr\"odinger operators,} 
J. Math. Soc. Japan {\bf 50} (1998), 359--377.

\bibitem{EI} P.~Exner and T.~Ichinose,
\emph{A product formula related to  quantum Zeno dynamics,}
Ann. H. Poincar\'e {\bf 6} (2005), 195--215.


\bibitem{GK} I.C.~Gohberg and M.G.~Krein,
{Introduction to the theory of linear non-selfadjoint operators
(in Hilbert space),}
Translations of Mathematical Monographs, Vol. 18, 
Amer. Math. Soc., Providence, R.I. 1969; its Russian edition,
Izdat. ``Nauka", Moscow 1965.

\bibitem{He} B. Helffer, 
\emph{Around the transfer operator and the Trotter--Kato formula,} 
Operator Theory: Advances and Appl. {\bf 79} (1995), 161--174.

\bibitem{Hi} F.~Hiai, 
\emph{Trace norm convergence of exponential product formula,} 
Lett. Math. Phys. {\bf 33} (1995), 147--158.

\bibitem{I}
T.~Ichinose, 
\emph{Time-sliced approximation to  path integral
and Lie--Trotter--Kato product formula,}
In: \lq\lq A Garden of Quanta",
Essays in Honor of Hiroshi Ezawa
for his seventieth birthday, World Scientific
2003, pp. 77--93.

\bibitem{INZ}
T.~Ichinose, H.~Neidhardt and V.~A.~Zagrebnov,
\emph{Trotter--Kato product formula and fractional powers of 
self-adjoint generators,}
J. Functional Analysis {\bf 207} (2004), 33--57;
\emph{Operator norm convergence of Trotter--Kato 
product formula,}
 Ukrainian Mathematics Congress---2001 (Ukrainian),  
100--106, Natsional. Akad. Nauk Ukraini, Inst. Mat.,  Kiev, 2002.  

\bibitem{ITak1} T.~Ichinose and S.~Takanobu,
\emph{Estimate of the difference between the Kac operator 
and the Schr\"odinger semigroup,} 
Commun. Math. Phys. {\bf 186}  (1997), 167--197.

\bibitem{ITak2} T.~Ichinose and S.~Takanobu,
\emph{The norm estimate of the difference between the Kac operator 
and the Schr\"odinger semigroup: A unified approach to the 
nonrelativistic and relativistic cases,}  
Nagoya Math. J. {\bf 149} (1998), 51--81.

\bibitem{ITak3} T.~Ichinose and S.~Takanobu,
\emph{The norm estimate of the difference between the Kac operator and 
the Schr\"odinger semigroup II: The general case including the 
relativistic case,} 
Electronic Journal of Probability {\bf 5} (2000), Paper 5, pages 1--47;
URL http://www.math.washington.edu/~ejpecp/

\bibitem{IT1}
T.~Ichinose and H.~Tamura, 
\emph{Error estimate in operator norm for Trotter--Kato product formula,}
{Integral Equations Operator Theory} {\bf 27} (1997), 195--207. 

\bibitem{IT2}
T.~Ichinose and H.~Tamura, 
\emph{Error bound in trace norm for Trotter--Kato product formula of 
Gibbs semigroups}, Asymptotic Anaysis {\bf 17} (1998), 239--266.


\bibitem{IT3}
T.~Ichinose and H.~Tamura, 
\emph{Error estimate in operator norm of exponential product formulas 
for propagators of parabolic evolution equations},
Osaka J. Math. {\bf 35} (1998), 751--770. 

\bibitem{IT4}
T.~Ichinose and H.~Tamura, 
\emph{The norm convergence of the Trotter--Kato product formula
with error bound,} 
Commun. Math. Phys. {\bf 217} (2001), 489--502;
{\sl Erratum}, ibid. {\bf 254} (2005),  255.


\bibitem{IT5}
T.~Ichinose and H.~Tamura,
\emph{Note on the norm convergence of the unitary Trotter product formula,}
Lett. Math. Phys.  \textbf{70}  (2004),  65--81.

\bibitem{IT6}
T.~Ichinose and H.~Tamura,
\emph{Sharp error bound on norm convergence of exponential 
product formula and approximation to kernels of Schr\"{o}dinger semigroups,} 
Comm. Partial  Differential  Equations \textbf{29} (2004), 1905--1918.

\bibitem{IT7}
T.~Ichinose and H.~Tamura, 
\emph{Exponential product approximation to integral 
kernel of Schr\"odinger semigroup
and to heat kernel of Dirichlet Laplacian,} 
J. Reine Angew. Math. \textbf{592} (2006), 157--188.

\bibitem{ITTZ}
T.~Ichinose, H.~Tamura, Hiroshi~Tamura and V.~A.~Zagrebnov,
\emph{Note on the paper 
\lq\lq The norm convergence of the Trotter--Kato product formula with
error bound" by Ichinose and Tamura,}
Commun. Math. Phys. {\bf 221} (2001), 499--510.

\bibitem{ITam}
T.~Ichinose and Hiroshi~Tamura, 
\emph{Imaginary-time path integral for a relativistic 
spinless particle in an electromagnetic field,} 
Commun. Math. Phys. {\bf 105}, 239--257(1986); For the final result 
of this see: 
T.~Ichinose, \emph{Some results on the relativistic Hamiltonian: 
Selfadjointness and imaginary-time path integral,} 
Differential Equations and Mathematical Physics, 
 pp. 102--116, International Press, Boston 1995.


\bibitem{Ka} T.~Kato, 
\emph{Trotter's product formula for an arbitrary pair of self-adjoint
contraction semigroups,} 
In: {Topics in Functional Analysis (essays dedicated to 
M. G. Krein on the occasion of his 70th birthday)}, edited by
I. Gohberg and M. Kac, {Academic Press, New York} 1978,
 pp.185--195. 

\bibitem{NZ0a} H.~Neidhardt and V.~A.~Zagrebnov, 
\emph{The Trotter--Kato product formula for Gibbs semigroups},
Commun. Math. Phys. {\bf 131} (1990), 333--346.


\bibitem{NZ0b} H.~Neidhardt and V.~A.~Zagrebnov, 
\emph{On the Trotter product formula for Gibbs semigroups},
Ann. Physik (7){\bf 47} (1990), 183--191.

\bibitem{NZ1} H.~Neidhardt and V.~A.~Zagrebnov, 
\emph{On error estimates for the Trotter--Kato product formula},
Lett. Math. Phys. {\bf 44} (1998), 169--186.

\bibitem{NZ2} H.~Neidhardt and V.~A.~Zagrebnov, 
\emph{Trotter--Kato product formula and operator-norm convergence},
Commun. Math. Phys. {\bf 205} (1999), 129--159. 

\bibitem{NZ3} H.~Neidhardt and V.~A.~Zagrebnov, 
\emph{Trotter--Kato product formula and symmetrically normed ideals},
J. Functional Analysis {\bf 167} (1999), 113--147. 

\bibitem{NZ4} H.~Neidhardt and V.~A.~Zagrebnov, 
\emph{Fractional powers of self-adjoint operators and Trotter--Kato product
 formula}, 
Integral Equations Operator Theory {\bf 35} (1999), 209--231. 


\bibitem{Ne} E.~Nelson, 
\emph{Feynman integrals and the Schr\"odinger equation,}
{J.\ Math. \ Phys.} {\bf 5} (1964), 332--343. 


\bibitem{Ro} Dzh.~L.~Rogava, 
\emph{Error bounds for Trotter--type formulas for self-adjoint operators,} 
Functional Analysis and Its Applications {\bf 27} (1993), 217--219.

\bibitem{S} R.~Schatten,
{Norm ideals of completely continuous operators,}
Springer, Berlin 1960.

\bibitem{Si} B.~ Simon, 
{Functional Integration and Quantum Physics,}
Academic Press, London 1979.


\bibitem{Su} M.~Suzuki, 
\emph{On the convergence of exponential operators
--- the Zassenhaus formula, BCH formula and systematic 
approximations,} 
Commun. Math. Phys. {\bf 57} (1977), 193--200.

\bibitem{Tak1} S.~Takanobu, 
\emph{On the estimate of the integral kernel for the Trotter 
product formula for Schr\"{o}dinger operators,} 
Ann. Probab. {\bf 25} (1997), 1895--1952.

\bibitem{Tak2} S.~Takanobu,
\emph{On the trace norm estimate of the Trotter product formula 
for Schr\"odinger operators,}
Osaka J. Math. {\bf 35} (1998), 659--682. 

\bibitem{Tam} Hiroshi Tamura, 
\emph{A remark on operator-norm convergence of Trotter--Kato product formula},
Integral Equations Operator Theory {\bf 37} (2000), 350--356.

\bibitem{Tr} H.~F.~Trotter, 
\emph{On the product of semigroups of operators,} 
Proc. Amer. Math. Soc. {\bf 10} (1959), 545--551.

\bibitem{V} S.~Varadarajan, Lie Groups, Lie Algebras, and
Their Representations, Springer, Berlin--Heidelberg--New York--Tokyo
1974, 1984.

\bibitem{Za1} V.~A.~Zagrebnov, 
\emph{The Trotter--Lie product formula for Gibbs semigroups},
J. Math. Phys. {\bf 29} (1988), 888--891.

\bibitem{Za2} V.~A.~Zagrebnov, 
{Topics in the Theory of Gibbs Semigroups,} 
Leuven Notes in Mathematical and Theoretical Physics,
Vol. 10, Leuven University Press 2003.

\end{thebibliography}
\end{document}